\documentclass[fleqn,10pt]{wlscirep}
\pdfoutput=1
\title{Anisotropic Friedel oscillations in graphene-like materials:
	The Dirac point approximation in wave-number dependent quantities revisited}

\author[1,2]{Tohid Farajollahpour}
\author[1,+]{Shirin Khamouei}
\author[3,+]{Shabnam Safari Shateri}
\author[1,2,*]{Arash Phirouznia}
\affil[1]{Department of Physics, Azarbaijan Shahid Madani
	University, 53714-161, Tabriz, Iran}
\affil[2]{Condensed Matter
	Computational Research Lab. Azarbaijan Shahid Madani University,
	53714-161, Tabriz, Iran}
\affil[3]{Department of Physics,University of Bonab, Bonab, East Azarbaijan, Iran }

\affil[*]{phirouznia@azaruniv.ac.ir}

\begin{abstract}
Friedel oscillations of the graphene-like materials are investigated theoretically 
for low and intermediate Fermi energies. Numerical calculations
have been performed within the random phase approximation. For intra-valley
transitions it was demonstrated that the contribution of the different Dirac
points in the wave-number dependent quantities, such as dielectric function
$\epsilon(\vec{q})$, has been determined by the orientation of the wave-number
with respect to the Dirac point position vector in $k$-space. Therefore identical
contribution of the different Dirac points is not automatically guaranteed by the
degeneracy of the Hamiltonian at these points. Meanwhile it was shown that the
contribution of the inter-valley transitions is always anisotropic even when the
Dirac points coincide with the Fermi level ($E_F=0$). This means that the Dirac
point approximation based studies give the correct physics only at high wave
length limit. The anisotropy of the static dielectric function reveals different
contribution of the each Dirac point. Additionally, the anisotropic $k$-space
dielectric function results in anisotropic Friedel oscillations in graphene-like
materials. Increasing the Rashba interaction strength slightly
modifies the Friedel oscillations in graphene-like materials. Therefore the
anisotropic dielectric function in $k$-space is the clear manifestation
of band anisotropy in the graphene-like systems. 
\end{abstract}
\begin{document}

\flushbottom
\maketitle
\begin{figure}[t]
	\includegraphics[width=0.5\linewidth]{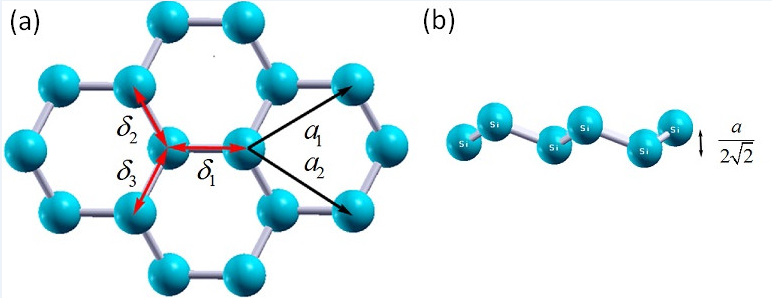}
	\centering
	\caption{(a) Hexagonal structure of buckled two dimensional lattice where ${\boldsymbol\delta}_{1}=a\left( 1,0,{1}/{2\sqrt{2}} \right)$, ${\boldsymbol\delta}_{2}=\frac{a}{2}\left(-1,{\sqrt{3}},{1}/{\sqrt{2}} \right)$, ${\boldsymbol\delta}_{3}=\frac{a}{2}\left( -1,{-\sqrt{3}},{1}/{\sqrt{2}} \right)$ are the nearest neighbors position vectors, the lattice vectors are ${{\boldsymbol a}_{1}}=\frac{a}{2}\left( 3,\sqrt{3} \right)$, ${{\boldsymbol a}_{2}}=\frac{a}{2}\left( 3,-\sqrt{3} \right)$, and the next nearest position vectors are ${{{\boldsymbol \delta }'}_{1}}=\pm {{\boldsymbol a}_{1}}$, ${{{\boldsymbol \delta }'}_{2}}=\pm {{\boldsymbol a}_{2}}$ and ${{{\boldsymbol \delta }'}_{3}}=\pm ({{\boldsymbol a}_{2}}-{{\boldsymbol a}_{1}})$. (b) Side view of buckled structure for silicene. \label{fig1}}
\end{figure}

%
%
\thispagestyle{empty}
At the present time two dimensional structures are one of the most
rich and fast growing fields of condensed matter physics.
Experimental observation of graphene in 2004 \cite{novoselov2004electric,geim2007rise}
has created a great motivation in scientists to the discovery and study
of the other possible two-dimensional (2D) allotropes of IV group
elements in periodic table such as silicene, germanene \cite{davila2016few} and
recently stanene \cite{xu2013large,zhu2015epitaxial}. These new 2D materials and other buckled
honeycomb lattice structures predicted in theoretical works \cite{guzman2007,cahangirov,liu2011,liu2013d+}
and several experimental synthesization have also been performed for realization
of these materials \cite{chen2012evidence,vogt2012silicene,chen2013spontaneous,feng2012evidence}.
The silicon and germanium analog of graphene with slightly buckled honeycomb
geometry were predicted to have a Dirac cone and the electrons follow the
massless Dirac equation near the Fermi level \cite{liu2011,cahangirov,guzman2007}.
Unlike graphene, the hybridization of $\pi$ bonds in silicene is not pure
and the structure of silicene shows a mixed hybridization. The $\pi$ electrons
in silicene are much more active than graphene and this lead to a different
structure from graphene \cite{lin2012much}. Similar to the graphene
structure, silicon atoms are arrayed in a hexagonal
lattice, but with a slight buckling that proved by first principle studies
which show low buckled silicene is thermally stable (Fig. \ref{fig1}) \cite{cahangirov}.
It was also shown that the electronic dispersion of the silicene near $K$
points of the first Brillouin zone is linear similar to the behavior of
Dirac materials \cite{guzman2007,cahangirov,takeda1994theoretical,durgun2005silicon}.
Since the flat configuration of silicene is not stable \cite{cahangirov}
the buckled configuration of silicene is more interesting for research activities. 
The spin orbit coupling (SOC) in silicene is more stronger than 
graphene which leads to relatively large energy gap at the Dirac points.
Strong SOC in silicene makes this monolayer a good candidate for topological
insulators and quantum spin Hall effect (QSHE) \cite{Ezawa2013,Tahir2013,liu2011,padilha2016new}.

It has been generally assumed that the degeneracy of the Dirac points provides
the identical contribution of these points in the physical quantities. This could
be a correct procedure and would be valid for calculation of the scalar quantities.
However, it should be noted that the Dirac points in $k$-space are not distributed
isotropically. This anisotropy has been dictated by the band anisotropy of the
honeycomb structures in the $k$-space. Therefore identical treatment of the Dirac
points automatically ignores anisotropy of the band energy. It seems that this
anisotropy could be appeared just at high Fermi energies. However, as it was shown in this work, even in the
case of low Fermi energies where the Fermi level could match the Dirac points the contribution of inter-valley
transitions are completely anisotropic.

Within the Dirac point approximation when the Dirac points have been treated
identically, anisotropic effects have been completely ignored at low Fermi
energies. Some of the anisotropic effects are raised by increasing the Fermi
energy up to the range of trigonal warping effects. However, even at low Fermi
energies, band anisotropy of the system manifests itself in the dielectric function 
$\epsilon(\vec{q})$, at least at the range of inter-valley transitions where
$q\sim\mid \textbf{K}_D-\textbf{K}_D'\mid$ in which $\textbf{K}_D$ and
$\textbf{K}_D'$ are different Dirac points. At the level of low Fermi energies, the band energy of the
system has been reduced to a cone-like dispersion. In this case the Fermi
level has been identified with a symmetric circle around the Dirac points
known as Fermi circle. Calculation of the wave number dependent
quantities should be performed with some care, since the direction of the
transferred momentum $\vec{q}$, determines the contribution of each Dirac
point in both intra-valley and inter-valley transitions.  For a given
transferred momentum $\vec{q}$, different Dirac points have not the same
contribution in this type of the physical quantities.This could be considered as another type of anisotropic
behaviors in wave number dependent quantities that originate from non-identical contribution
of the Dirac points. This work attempts
to provide some insight into the limitations of identical treatment of the
Dirac points. Results of the current work emphasize the need for a systematic
revision of identical treatment of the Dirac points in
different types of quantities. Specifically, we analyze the robustness of
the band anisotropy in graphene and other honeycomb systems which manifests
itself in the dielectric function and Friedel oscillations of the system.
To this end, random phase approximation (RPA) \cite{mahan2000many} is employed beyond the Dirac
point approximation. In this case at low Fermi energies, 
validity of the  identical contribution of the Dirac points 
in the dielectric function could be examined within this approach.

It is obvious that the contribution of the nonlinear part of the
energy dispersion at high Fermi energies could be given beyond
the Dirac point approximation. If we use the same approach at the level of low Fermi energies. This ensures us that the correct contribution of each Dirac point has been taken into account. 
It can be shown that even at low Fermi energies the band
induced anisotropy could be observed in graphene-like materials. Meanwhile the anisotropic effects which have been observed at low Fermi energies have nothing to do with the nonlinear part of the energy dispersion which is available beyond the Dirac point approximation. This could be understood if we consider that band anisotropy is present both at high and low energy limits. 
The unique optical and electronic properties of graphene-like systems such as silicene
has made these materials a good candidate for plasmonics applications.  Meanwhile
plasmonic-based studies has already been performed for graphene however, the other
graphene-like systems are known as highly appealing subjects for this field of
condensed matter physics \cite{grigorenko2012graphene,bonaccorso2010graphene,tabert2014dynamical,PhysRevB.88.115420}.

\begin{table}
	\centering
	\caption{Lattice constant and Energy scales for graphene and other buckled honeycomb materials \cite{PhysRevB.84.195430,ezawa2012,liu2011,Min}.
	\label{table-structure}}
	\begin{tabular}{crrrr}
		\hline
		\hline
		\ \ material \ \  & \ \ $a$\ \  & \ \ $t$\ \  & \ \ $t_{SO}$ \ \  & \ \ $t_{intR}$\\
		\hline\noalign{\smallskip}
		silicene           &      3.86 ~\AA   & \ \  1.6 eV    & \ \   0.75 meV & \ \ 0.46 meV\\
		\\
		germanene          &     4.02 ~\AA   & \ \  1.3 eV  & \ \  8.27 meV & \ \ 7.13 meV \\
		\\
		graphene           &    2.46 ~\AA    & \ \   2.8 eV  & \ \ 0.00114 meV       & ---    \\
		
		\hline
		\hline
	\end{tabular}
\end{table}

Friedel oscillation has been reported for graphene in low energy Hamiltonian
which relies on the Dirac-cone approximation
\cite{scholz2012dielectric,wunsch2006dynamical,gomez2011measurable,Cheianov}. It is important to note that the information about the possible topological phase transitions could be captured by Friedel oscillations. Results of the Friedel oscillations in silicene demonstrates that there is a connection between the Friedel oscillations and topological phase transition\cite{ChangZhang}.
These approaches ignore the contribution comes from the nonlinear part of the
spectrum and can be reasonable when the Fermi energy is close
to the Dirac points. Nevertheless, as mentioned before identical treatment of the Dirac points within the Dirac
point approximation cannot capture the anisotropic contribution
of these points in wave-number dependent quantities. In this work calculations have been performed beyond the Dirac point
approximation in which all possible types of the band anisotropy, including the nonlinear and linear
parts of the spectrum, could be considered.  Meanwhile, it should be noted that the linear
dispersion at Dirac points and even existence of Dirac cones in silicene is being seriously debated\cite{schmidt2013effective,Quhe2014}.
Motivated by the mentioned points, we have
performed current numerical study to obtain a better understanding  about the limitations of
the Dirac point approximation in graphene-like systems.\\

\section*{Theoretical framework}

Graphene-like materials could be considered as honeycomb lattice structures, 
similar to graphene meanwhile the SOC of buckled honeycomb structures
contains parallel and perpendicular terms. The Hamiltonian of the buckled 
honeycomb lattice in tight-binding approximation in the presence of SOCs 
can be written as 
\begin{eqnarray}
H=-t\sum\limits_{\left\langle ij \right\rangle \alpha }{c_{i\alpha }^{\dagger }}{{c}_{j\alpha }}+
i{{t}_{SO}}\sum\limits_{\left\langle \left\langle ij \right\rangle  \right\rangle \alpha \beta }{{{u}_{ij}}c_{i\alpha }^{\dagger }\sigma _{\alpha \beta }^{z}{{c}_{j\beta }}}
-i{{t}_{intR}}\sum\limits_{\left\langle \left\langle ij \right\rangle  \right\rangle \alpha \beta }{{{\mu }_{ij}}\hat{c}_{i\alpha }^{\dagger }\left( \vec{\sigma }\times {{{\vec{d}}}_{ij}} \right)_{\alpha \beta }^{z}{{{\hat{c}}}_{j\beta }}}+
i{{t}_{extR}}\sum\limits_{\left\langle ij \right\rangle \alpha \beta }{\hat{c}_{i\alpha }^{\dagger }\left( \vec{\sigma }\times {{{\vec{d}}}_{ij}} \right)_{\alpha \beta }^{z}{{{\hat{c}}}_{j\beta}}}
\end{eqnarray}
where the operator ${\mathop{c^\dagger_{j\alpha}}} (\mathop{c}_{j\alpha})$
creates (annihilates) an electron with spin $\alpha$ at site $j$ and $t$ is
the nearest neighbor hopping amplitude. The values of these parameters for
different materials are given in table~ \ref{table-structure}. The ${t}_{SO}$ is the next-nearest neighbor hopping, ${{u}_{ij}}={{{{\vec{d}}}_{i}}\times {{{\vec{d}}}_{j}}} /{\left| {{{\vec{d}}}_{i}}\times {{{\vec{d}}}_{j}} \right|}$ where ${{\vec{d}}}_{i}$
and ${{\vec{d}}}_{j}$ are the two nearest bonds that connect the next-nearest
neighbors, Where ${{u}_{ij}}=1$ if the next-nearest neighbor hopping is
counterclockwise and ${{u}_{ij}}=-1$ when it is clockwise with respect to
the positive $z$ axis \cite{ezawa2013spin}. The $ {\left\langle \left\langle ij \right\rangle  \right\rangle } $
run over all the next-nearest neighbor hopping sites and ${\sigma}_{z}$ is
the Pauli matrix. ${t}_{\operatorname{int}R}$ and $t_{ext R}$ are the 
strength of intrinsic and extrinsic Rashba 
SOCs respectively and $\mu_{ij}=+1(-1)$ stands for the A (B) site. 
The strength of the external Rashba coupling can be manipulated
by an external gate voltage or by the selected substrate. The extrinsic
Rashba coupling arises as a result of the inversion symmetry
breaking due to an applied perpendicular electric field or
interaction with substrate \cite{bychkov1984oscillatory}.

Dielectric function and the screening of the charged impurity
and also the dynamical polarization which gives collective
excitations could be captured by the polarization function
$\Pi(\omega,q)$. Dielectric function and collective density 
oscillations of an electron liquid (plasmons), have been 
observed in different metals and superconductors
\cite{ando1982electronic,giuliani2005quantum}. At the static limit
($\hbar \omega=0$) polarization function gives the screening behavior
of the coulomb potential. The dielectric function is relevant to plasmonic
studies meanwhile the transport and phonon spectra are also another
relevant fields \cite{kaasbjerg2013acoustic}.
\begin{figure}[h]
	\includegraphics[width=0.35\linewidth]{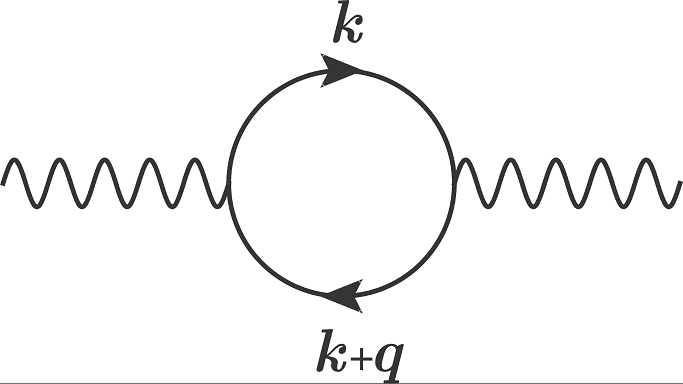}
	\centering
	\caption{The bare polarization bubble diagram corresponding
		to Eq. (\ref{aa}). \label{bubble}}
\end{figure}
The electron-electron interaction has been
considered within the random phase approximation characterizes by the
density-density correlation function or polarization function (Fig. \ref{bubble})
\cite{pyatkovski2008polarization,pyatkovskiy2009dynamical,
	scholz2012dielectric,gorbar2002magnetic,hwang2007dielectric,wunsch2006dynamical,
	sensarma2010dynamic,scholz2013plasmons,giuliani2005quantum,mahan2000many}.
In this approach dielectric function is given by
\begin{equation}\label{bb}
\epsilon \left( \omega ,\vec{q} \right)=1-V\left( q \right)\Pi \left( \omega ,\vec{q} \right)
\end{equation}
Where $V\left( q \right)$ is the 2D Coulomb potential, here  $V\left( q \right)={2\pi e^2 }/{q}$.
Within the Dirac point approximation an effective Coulomb potential could be employed in which
$V\left( q \right)={2\pi \alpha }/{q}$ and $\alpha$ is the ratio of coulomb to kinetic energy
and named effective fine structure constant where equal to $\alpha ={{{e}^{2}}}/{\left( \hbar {{\varepsilon }_{0}}\upsilon  \right)}$ and ${{\varepsilon}_{0}}$ is the bare dielectric constant. 
Unlike to the graphene where the value of fine structure constant could be
determined experimentally in different substrates \cite{hwang2012fermi},
for other buckled honeycomb lattices one can set $\alpha=0.8$ \cite{tabert2014dynamical}.
The polarization function in one loop approximation is calculated directly from
the bubble diagram that shown in Fig (\ref{bubble}).
\begin{eqnarray}\label{aa}
\Pi \left(\omega, \vec{q}  \right)&=&\sum\limits_{s {s}' k k'}{\frac{f_{k}^{s}-f_{k'}^{{{s}'}}}{\omega +E_{k}^{s}-E_{k'}^{{{s}'}}} |<k' \lambda^{s'}_{k'}\mid e^{iq.r}\mid k \lambda^s_{k}>|^2}\nonumber\\
&=&\sum\limits_{s{s}'k}{\frac{f_{k}^{s}-f_{k+q}^{{{s}'}}}{\omega +E_{k}^{s}-E_{k+q}^{{{s}'}}}{{F}_{s'{s}}}\left( \vec{k}+\vec{q},\vec{k} \right)},
\end{eqnarray}
here, the summation performed over the full Brillouin zone
and all of the spin and pseudo-spin dependent eigenstates
in which $f_{k}^{s}=\frac{1}{\exp \beta \left( E_{k}^{s}-E_F  \right)+1}$
is the Fermi distribution function, and $ E_F$  is the Fermi energy.
The form factor is given by $F_{s',s}( \vec{k}',\vec{k} )=|<k' \lambda^{s'}_{k'}\mid e^{iq.r}\mid k \lambda^s_{k}>|^2=|<k' \lambda^{s'}_{k'}\mid k \lambda^{s}_{k}>|^2\delta_{\vec{k}',\vec{k}+\vec{q}}$
in which $\mid k \lambda^{s}_{k}>=\mid k>\otimes \mid \lambda^{s}_{k}>$
are the eigenstates of the Hamiltonian where $\mid \lambda^{s}_{k}>$ is
the eigenstate in the spin and pseudo-spin subspaces. $\delta_{\vec{k}',\vec{k}+\vec{q}}$
represents the momentum conservation rule as a general condition for contributing transitions.

Within the Dirac point approximation the above expression of the polarization function
has been assumed to be \cite{wunsch2006dynamical}
\begin{eqnarray}
\Pi \left( \omega ,\vec{q}  \right)=g\sum\limits_{s{s}'k}{\frac{f_{k}^{s}-f_{k+q}^{{{s}'}}}{\omega +E_{k}^{s}-E_{k+q}^{{{s}'}}}{{F}_{{s}'s}}\left( \vec{k}+\vec{q},\vec{k}  \right)}\label{Pd}
\end{eqnarray}
where $g$ is valley degeneracy factor and the summation runs around a single dirac point.
Moreover in the absence of the spin-orbit interactions the form factor is reduced to:
${{F}_{s'{s}}}\left(\vec{k}+\vec{q},\vec{k} \right)=\frac{1}{2}[1+s{s}'cos\left( 2\theta  \right)]$,
with $ \theta$  being the angle between $k$ and $k+q$. It should be noted that in this
relation the degeneracy factor, $g$, implies identical contribution of the different Dirac points
in the polarization function at given $\vec{q}$. As discussed before the valley degeneracy could
result in identical contribution of the Dirac points in scalar quantities such as total energy.
Nevertheless this degeneracy cannot indicate the identical contribution of the Dirac points in
the wave number dependent quantities such as polarization function.
Consequently, this type of calculations should be performed beyond the Dirac point
approximation even when the Fermi energy level lies close to the cones intersecting points.
This is due to the fact that the contribution of the different Dirac points, $\vec{K}_D$
and $\vec{K}_D'$, is not the same and depends on the
location of the Dirac point in the $k$-space that has been identified by $\vec{K}_D$.

When the integration has been reduced to the Fermi
circle of a single Dirac point this assumption automatically ignores the contribution
of the inter-valley transitions in which the initial and final states belong to different
Fermi circles. This could take place when the transferred momentum, $q$, satisfies
$q\sim\mid\vec{K}_D-\vec{K}_D'\mid$. This means that the general form of the Dirac
point approximation ignores the inter-valley transition. The anisotropy of the dielectric
function results from this type of transitions when the Fermi energy is exactly zero. 

Within a second-order perturbation approach it has been realized that the exchange interaction of the localized spins $S_1$ and $S_2$ with the conducting electrons results in an effective magnetic interaction between these localized magnetic moments known as RKKY interaction given by\cite{hwang2008screening} $H_{RKKY}(r)=JS_1.S_2\Pi(r)$  in which $J$ is the exchange coupling constant between the conducting electrons and localized magnetic moments and $\Pi(r)$ is the Fourier transform of the $k$-space polarization function $\Pi(q)$ . Therefore the characteristic properties of this interaction  could be captured by the polarization function of the mediating electrons. Accordingly it is expected that the anisotropy of the polarization function could manifest itself in the anisotropy of the RKKY interaction as it appeared in the dielectric function.

The polarization function could be separated into the inter-band (if $s\ne {s}'$ ) and
the intra-band (if $s=s'$) contributions\cite{sensarma2010dynamic}. In addition each of
these contributions could be classified as intra-valley and inter-valley contributions
that correspond to different transitions in which the transferred momentum, $\vec{q}$,
is $q\leq k_F<\mid \textbf{K}_D-\textbf{K}_D'\mid$ or $q\sim \mid \textbf{K}_D-\textbf{K}_D'\mid$
respectively (where $k_F$ is the radius of the Fermi circle and $\textbf{K}_D$, $\textbf{K}_D'$
are different Dirac points). To illustrate the oscillations of the induced charge density a charged impurity
has been considered to be inserted in the honeycomb structure. The static polarization function is of particular importance as it determines the screened
potential of a charge impurity. The screening particle density ${\delta n}(r)$ due
to the central impurity $Ze$ is,
\begin{equation}
{{ \delta n}}\left( \vec{r} \right)=Ze\frac{1}{{{\left( 2\pi  \right)}^{2}}}\int{\left[ \frac{V\left( \vec{q} \right)\Pi \left( 0 ,\vec{q} \right)}{\epsilon \left(0, \vec{q} \right)} \right]\exp \left( i\vec{q}.\vec{r} \right)d^2q}.
\end{equation}
It is really important to note that the above integration goes beyond the limit of intra-valley transitions (the radius of Fermi circle) and therefore the contribution of the inter-valley transitions should be included. This means that the pattern of Friedel oscillations requires the whole information of the static response function in the first Brillouin zone. Therefore inter-valley transitions between the different Dirac cones should be included. Therefore this type of transitions cannot be captured within the single Dirac cone approximation.

We want to remark that the Friedel oscillation curves in the RPA,
Hubbard vertex correction and Singwer-Sj{\"o}lander are very similar\cite{hubbard1958description,singwi1970electron,mahan2000many}.
The reason, for example in the Hubbard Model, is that the Hubbard local
field factor ${{G}_{H}}(q)$ is appeared in correlations that are deal
with two-body or more, however the screening is essentially a one-body
property and the correlations are not appear in one-body amplitudes
\cite{mahan2000many}.
\section*{Non-identical contribution of different Dirac cones}
\label{app}

\begin{figure}[t]
	\includegraphics[width=0.35\linewidth]{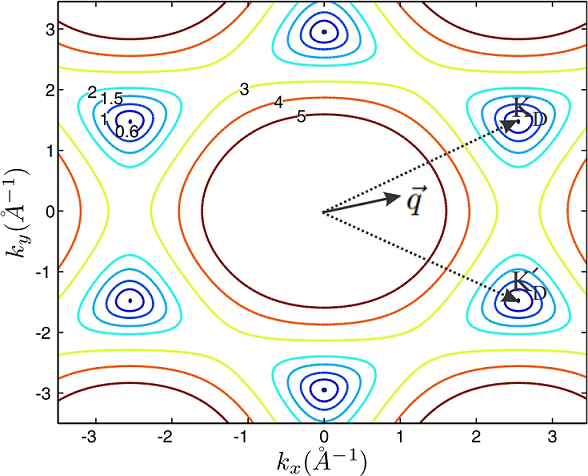}
	\centering
	\caption{Dirac points of monolayer graphene and Fermi curves
		at different Fermi energies. Fermi contours have been depicted for
		$E_F=0.6eV,1.0eV~1.5eV,~2.0eV,~3.0eV,~4.0eV$ and $5.0eV$. For a given
		wave vector $\vec{q}$ the contribution of the different Dirac points
		on $\epsilon(\vec{q})$ strictly depends on the orientation and position
		of the $\vec{q}$ with respect to the six Dirac vectors. Trigonal warping
		of the Fermi curves at different Fermi energies has also indicated in this
		figure. Single Dirac cone approximation could take into account the
		anisotropic effects comes from the trigonal warping of a single Fermi
		curve, however, since the orientation of the deformed Fermi curves are
		not the same, the anisotropic contribution of the other cones are not identical.\label{Dirac1}}
\end{figure}

As depicted in Fig.~(\ref{Dirac1}) at low Fermi energies the Fermi curves
have been appeared as separated  islands around each  Dirac point. Therefore
the amount of the dielectric and polarization functions in a given $\vec{q}$
wave number have significantly been determined by the orientation of the
wave number with respect to the Dirac points position vectors. This anisotropy
of the $q$-space is reflected in the real space quantities such as
Friedel oscillations.   

When the Fermi level is appeared as distinct
circular curves in k-space. One might conclude that
the circular shape of the Fermi contour around of the Dirac points implies
that all of the extractable physical properties of the system should be
isotropic as long as the Dirac point approximation is applied. However
it should be considered that the isotropic and circular shape of the Fermi
contours around each of the Dirac points cannot results in isotropic properties at least in when the inter-valley transitions are taken into account. This is due to the anisotropy of the band structure in honeycomb systems which indicates that the Dirac points
themselves are located in k-space  in an anisotropic manner. Accordingly it is important to note that the results of present study cannot be compared with the results that have been obtained within the single Dirac point approximation\cite{Deng2015}.

Meanwhile, for calculation
of vector and tensor dependent quantities we should consider that the
contribution of inter-valley transitions cannot be identical in all of the
$q$-space directions even when $E_F=0$. Moreover when $E_F>0$ both of the
intra-valley and inter-valley transitions result in anisotropic dependence
of the dielectric function in $q$-space. Therefore different directions
in of a given momentum transfer, $q$, have not identical contribution even when the Dirac points
are degenerate. Accordingly the conventional Dirac point
approximation could not describe all of the physics of the vector or
vector dependent parameters at low wave-length limit. This could also result in anisotropic
electric and thermal conductivity in graphene-like
materials for short range scatterers (for example in the case of the delta-shaped scatterers) in
which all of the intra-valley and inter-valley scatterings are possible.

When the Dirac points are not located exactly on the Fermi level the intra-valley transitions could take place within the range of $q\leq k_F$.
Meanwhile we have assumed that the Fermi energy is still low enough to employ the linear
dispersion relation of the Dirac cone. It can be shown that both of
the intra-valley ($q\leq k_F$ for this case) and inter-valley transitions
should be considered as anisotropic contributions in the dielectric function.
In this case since the inter-band transitions ( $\mid k \lambda^s_k>\rightarrow\mid k' \lambda^{s'}_{k'}> ~~ s\neq s' $ )
are absent in the static limit ($\hbar\omega=0$) therefore all of the
contributing terms (both intra-valley and inter-valley transitions) are
intra-band. Consequently, the contribution of $q\sim 0$ transitions in the
static dielectric function decreases by increasing the Fermi energy. Since it can be shown that for $E_F\neq 0$
we have $F_{s s'}\left( \vec{k}+\vec{q},\vec{k}\right)=0~~$ when $q=0$
and $ s\neq s'$.

\begin{figure}[t]
	\includegraphics[width=1.0\linewidth]{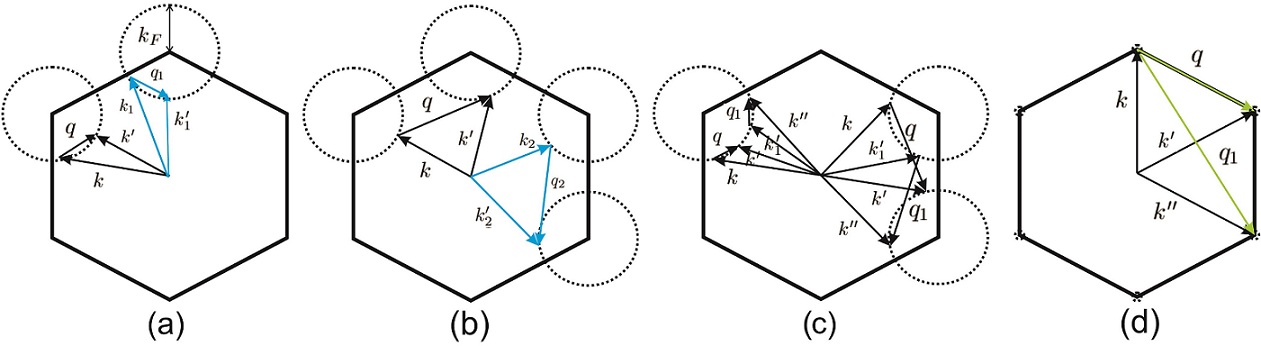}
	\centering
	\caption{(Color online) Intra-valley (a) and inter-valley (b) transitions for a given
		transferred momentum $\vec{q}$. Dashed circles indicate the Fermi circles of the honeycomb system in the absence of the spin-orbit couplings. When the momentum conservation rule
		is satisfied for $\vec{q}$ the initial and final states should be
		placed on the Fermi circles. In this case the contribution of the
		given states (black vectors) is identical with the contribution of
		the sixfold-rotated states (cyan vectors). One can imagine about
		another type of possible transitions (c) with constant value of the
		transferred momentum $q=q_1$ between the equi-energy states
		$E^s_k=E^s_{k'}=E^s_{k'_1}=E^s_{k''}$ where the corresponding pair vectors
		($\vec{q}$  $\vec{q}_1$), ($\vec{k}$ $\vec{k'}_1$) and ($\vec{k'}$ $\vec{k''}$)
		are not related by sixfold symmetry operators e.g. $\mathcal{R}^n_{{2\pi}/{6}}\vec{q}\neq\vec{q}_1$.
		It can be shown that form factor of these transitions are different i.e. $F(k,k')\neq F(k'_1,k'')$.
		At zero Fermi energy (d) i.e. when $k_F=0$ intra-valley transitions occur at $q=0$
		which result in central peak of the dielectric function. However inter-valley
		transitions (light green vectors) are still the source of anisotropy of the
		dielectric function.\label{qval}}
\end{figure}

Different types of transitions could be contributed in the dielectric
function of the honeycomb structures. In this case transitions could
be either intra-valley or inter-valley. Where in the intra-valley
transitions initial and final states $\vec{k}$ and $\vec{k'}$ belong
to the same Dirac valley cone while in the inter-valley transitions
$\vec{k}$ and $\vec{k'}$ belong to different Dirac cones (Fig. \ref{qval}
(a) and (b)). The momentum conservation rule for each transition between
the states $\vec{k}$ and $\vec{k'}$ with transferred momentum $\vec{q}$
could be satisfied when $\vec{k}$ and $\vec{k'}$ sweep the Fermi circles
as shown in Fig. \ref{qval}.  It can be inferred that this condition could
be satisfied for intra-valley transitions when $0 \leq  q  \leq  k_F$ where
$k_F$ is the radius of the Fermi circle Fig.~\ref{qval} (a) and inter-valley
transitions could take place $q\sim \mid \textbf{K}_D-\textbf{K}_D'\mid$ where
$\textbf{K}_D$ and $\textbf{K}_D'$ are different Dirac points.

Fig. \ref{qval} shows the Fermi circles of a planar honeycomb lattice has
been depicted in the absence of the SOCs. It can be shown
that due to this six-fold band rotational symmetry of the system if the
transition rule is satisfied for a given transferred momentum ($\vec{q}$)
it will also be satisfied for the sixfold rotated wave number
$\mathcal{R}^n_{{2\pi}/{6}} \vec{q}$ (Fig. \ref{qval} (a) and (b)).
In which $\mathcal{R}_{{2\pi}/{6}}$ is the sixfold rotation operator. This means that
\begin{eqnarray}\label{rule}
\delta_{\vec{k}+\vec{q},\vec{k'}}&=&\delta_{\mathcal{R}^n_{{2\pi}/{6}}\vec{k}+\mathcal{R}^n_{{2\pi}/{6}}\vec{q},~\mathcal{R}^n_{{2\pi}/{6}}\vec{k'}}\nonumber\\
&=&\delta_{\vec{k}_n+\vec{q}_n,\vec{k'}_n}
\end{eqnarray}

Meanwhile the form factor of the graphene is also invariant under
the sixfold rotations in the absence of the spin-orbit couplings.
\begin{eqnarray}\label{form2}
{{F}_{s'{s}}}\left( \vec{k}+\vec{q}, \vec{k}\right)={{F}_{{s}'s}}\left(\mathcal{R}^n_{{2\pi}/{6}}\vec{k}+\mathcal{R}^n_{{2\pi}/{6}}\vec{q}, \mathcal{R}^n_{{2\pi}/{6}}\vec{k} \right).
\end{eqnarray}

In both cases i.e. for inter-valley and intra-valley transitions
band symmetry of the honeycomb structures manifests itself as
$E_{\vec{k}}^{s}=E_{\mathcal{R}^n_{{2\pi}/{6}}\vec{k}}^{s}$.
Therefore Eq. \ref{aa} reveals that for a given transferred momentum,
$\vec{q}$, satisfying the transition rule $\vec{k}=\vec{k}'-\vec{q}$.
The contribution of the $\vec{k}$-state in the dielectric function is
identical with the contributions of the rotated states:
$\mathcal{R}_{{2\pi}/{6}}\vec{k}$,  $\mathcal{R}^2_{{2\pi}/{6}}\vec{k}$,...
$\mathcal{R}^6_{{2\pi}/{6}}\vec{k}$ regardless of the inter-valley or
intra-valley nature of the transitions. In the other words it could be
inferred that $\epsilon(\vec{ q})$ $=\epsilon( \mathcal{R}_{{2\pi}/{6}} \vec{ q})$ $=\epsilon( \mathcal{R}^2_{{2\pi}/{6}} \vec{ q})$...$=\epsilon( \mathcal{R}^5_{{2\pi}/{6}} \vec{ q})$.
For example in the case of intra-valley transitions the different
Dirac point or Fermi curves which have been related by sixfold
rotation operators have the same contribution in the dielectric
function for those transferred momentums which have the same
symmetry relation. Consequently, the contribution of the Fermi circle
located around the $\vec{K_D}$ Dirac point on the dielectric function
of $\vec{q}$  is identical with the contribution of $\vec{K}_{D}^{(i)}$
Fermi circle on the dielectric function of of $ \mathcal{R}^i_{{2\pi}/{6}} \vec{ q}$
where these two Dirac points are related by $\vec{K}_{D}^{(i)}=\mathcal{R}^i_{{2\pi}/{6}} \vec{K_D} $.
If we continue the same procedure for parallel wave numbers,
which satisfying the transition rule,
one can realize the anisotropy of the dielectric function
which manifests itself by sixfold symmetric curve at low
transferred momentums Fig.~\ref{eps-mg-ef-05}.
These relations identify different class of the states which
have identical contribution in the dielectric function i.e.
the for different states with transferred momentum $q$  in this case
the class of the identical contributions for both inter and intra
valley transitions is specified by
\begin{eqnarray}\label{class}
[q]=\{\vec{q},~ \mathcal{R}_{{2\pi}/{6}}\vec{q},~ \mathcal{R}^2_{{2\pi}/{6}}\vec{q},~ ...\mathcal{R}^5_{{2\pi}/{6}}\vec{q}\}.
\end{eqnarray}
Which corresponds to the transitions: $\vec{k}\rightarrow\vec{k}+\vec{q}$, $\mathcal{R}^5_{{2\pi}/{6}}\vec{k}\rightarrow\mathcal{R}_{{2\pi}/{6}}\vec{k}+\mathcal{R}_{{2\pi}/{6}}\vec{q}$,
...$\mathcal{R}^5_{{2\pi}/{6}}\vec{k}\rightarrow\mathcal{R}_{{2\pi}/{6}}\vec{k}+\mathcal{R}_{{2\pi}/{6}}\vec{q}$.

The first consequence of the above argument is that the different Fermi curves
of each Dirac point have not identical contribution on the dielectric function
at a given $\vec{q}$  wave number. In the other words the contribution of each
Fermi circle (corresponding to $K_D$ dirac point) with a given transferred momentum,
$\vec{q}$, has been determined by the orientation of the $\vec{q}$ with respect to
the $K_D$.  When $\vec{q}$ satisfies the transition rule for a specific Dirac cone
(for example in an intra-valley process) this rule will be satisfied for
$\mathcal{R}^n_{{2\pi}/{6}}\vec{q}~~(n=1..5)$ at other Dirac cones and $\vec{q}$
itself could not satisfy the momentum conservation rule or could not give the same
contribution at these Dirac cones. Therefor the contribution of different Fermi
circles on the dielectric function of $\vec{q}$ is not identical. Accordingly
the dielectric function should be anisotropic in the $q-$space with sixfold
symmetry which was originated from the symmetry of the band structure.

All of the other possible transitions with a given fixed value of the
transferred momentum could take place between the isoenergy states as
shown in Fig. \ref{qval} (c). In this group of the transitions, transferred
momentum is the same $q=q_1$ and both of the initial final states are
located at Fermi circle $E^s_k=E^s_{k'}=E^s_{k'_1}=E^s_{k''}=E_F$. However,
corresponding pair vectors are not related by sixfold symmetry operators
i.e. $\mathcal{R}^n_{{2\pi}/{6}}\vec{q}\neq\vec{q}_1$, $\mathcal{R}^n_{{2\pi}/{6}}\vec{k}\neq\vec{k}'_1$
and $\mathcal{R}^n_{{2\pi}/{6}}\vec{k'}\neq\vec{k}''$. The transition rule has
been satisfied for these transition where we have $\vec{k}+\vec{q}=\vec{k}'$
and $\vec{k}'_1+\vec{q}_1=\vec{k}''$. Meanwhile, the form factor of the
transitions are not the same $F(k,k')\neq F(k'_1,k'')$ which implies that
the corresponding contributions are not identical (Fig. \ref{qval} (c)).

When the Fermi energy located at Dirac points i.e. $E_F=0$ then the intra-valley
transitions occur in $\vec{q}=0$ between different bands (Fig \ref{qval} (d)).
Unlike the previous case the intra-valley contributions are identical here.
These contributions result in the central peak of the dielectric function.
Since the $V(q)=2\pi e^2/q$ diverges at $q=0$ where the intra-valley
transitions could contribute at this point. However the inter-valley
transitions occur away from the $\Gamma$-point ($q=0$) result in anisotropic
contributions as the previous case. Therefore one can expect that the
anisotropy of the dielectric function would be appeared at
$q \geq \mid \textbf{K}_D-\textbf{K}_D'\mid$ Fig. (\ref{eps-mg-0-2d}).
\section*{Discussions and numerical results}
In conclusion, we have analyzed the band induced anisotropic effects
in graphene-like structures.
We have also investigated the influence of the spin-orbit
interactions on dielectric function and Friedel oscillations
in this type of materials.

Due to the foregoing discussions for wave number-dependent or non-scalar quantities
such as dielectric function, electric and thermal conductivities we have to concern about
the position of the Dirac points relative to the direction of the characteristic vector
of the physical quantity (such as transfered momentum) even when the Dirac point approximation is valid. For sharp scattering potentials we have to consider the inter-valley transitions in this type of the quantities. Within the Dirac point approximation the integration over the state-resolved contributions is generally performed over a single Fermi circle. This could be a correct approach when we assume the identical
contribution of each Dirac point and ignore the inter-valley transitions.
In this case
if we put aside the Dirac point approximation and perform
the integration over the whole Brillouin zone the correct contribution of each Dirac
point could be obtained. This means that beyond the Dirac point approximation dielectric function and therefore
Friedel oscillations for graphene-like honeycomb structures should be anisotropic in
k-space and the real space respectively.


Based on RPA formalism, which accounts
for electron-electron scattering, we have shown that the Dirac
points have not identical contribution for wave number dependent
quantities such as dielectric function even when the Fermi energy
is close to these points.

The main limitations for the use of the
Dirac point approximation has been discussed within the current
work. Non-identical contribution of the Dirac points results in
anisotropic dielectric function in $k$-space. Moreover, the anisotropy of the dielectric function leads to anisotropic Friedel oscillation in graphene-like materials.
We have shown that the isotropic Friedel oscillations cannot be considered as a consequence of the Dirac point approximation itself. Since in this case the isotropic Friedel oscillations could be obtained just when the inter-valley transitions are ignored.

The influence of the Rashba SOC on the dielectric function
and Friedel oscillations has also been discussed in the present study
where we have shown that increasing the Rashba coupling strength cannot results in significant change in the dielectric function and Friedel oscillation. Meanwhile in the presence of the
Rashba interaction the inversion symmetry of the dielectric function
has been lifted in $k$-space. The extrinsic Rashba
interaction could be manipulated by an external gate voltage. By
changing the extrinsic Rashba coupling strength, polarization
function and consequently the dielectric function will be altered. 
Dielectric function has been determined by density-density correlations
in the system. In the present work this correlation function has been
obtained within the random phase approximation. Results of the numerical
calculations depicted in the following figures of this section.

As mentioned before the band induced anisotropic effects could be obtained
even when the Fermi energy is close to the Dirac points.
This is due to the fact that the optical transitions which could be available
for a given transferred momentum $\vec{q}$ are not identical for different Dirac
points in $k$-space. At the level of the single Dirac point approximation anisotropic
position of the Dirac points and its relevant effects have been ignored. Meanwhile,
the anisotropy of single Fermi circle reshaping, which arises by increasing the
Fermi energy, could be obtained within the single Dirac cone approximation. In the present work we have
obtained the anisotropic properties which could be originated from the anisotropic
location of the Dirac points at low Fermi energies.

According to the numerical results, two-dimensional graphene like materials show
anisotropic Friedel oscillations beyond the Dirac cone approximation even when
the Dirac points located at the Fermi level. In addition results of the present
study show that the extrinsic Rashba coupling has not any considerable effect on
the Friedel oscillations. Meanwhile the influence of the spin-orbit couplings is relatively evident
in the Friedel oscillations of the monolayer germanene.

It is important to consider that the general integral expression for the polarization
function (Eq. \ref{aa}) goes beyond the states in which the Dirac point approximation
is not valid. Nevertheless, far from the Dirac points most of
the states in each band are empty with no contribution in the dielectric function in the static limit ($\hbar \omega =0$) even
at room temperature.
For gap-less structures, at room temperature, thermal transitions could be taken
place within the range of thermal energy $K_BT=0.025$eV. Therefore, when the Fermi
level is located at the Dirac point, the contribution of the states which were not
in the permitted range of the Dirac point approximation automatically have been
eliminated by the distribution function implemented in Eq. (\ref{aa}).

The linear dispersion relation (and therefore circle like Fermi curves)
around the Dirac points valid even up to $E_F \sim 1$ eV in graphene and the thermal
transitions at room temperature with $K_BT=0.025$ eV could not induce any considerable
contribution from those states which have been located far from the Dirac points.
So it seems that the Dirac point approximation could still describe the
physics of the honeycomb lattice and the linear dispersion relation around the Dirac
points could be employed for the calculation of the dielectric function. This means
that the non-linear part of the band structure and the anisotropy that might be induced by this part could be ignored. This is due to the fact that this part of
the band structure (which could be considered the energy states with $E^s_{k}>1 eV$)
has not been occupied even as a result of the thermal transitions or could not contribute in the isoenergy transitions of the static limit. However, in the
current work we have shown that the contribution of the Dirac points in wave-number
dependent quantities should be calculated with some care even when the Dirac point
approximation is valid.

It seems that the degeneracy of the K-points results in identical contribution of each Dirac
point in all of the physical quantities when the Fermi energy is close to the Dirac points.
However, it should be noted that, this is not the case for some of the quantities which directly
depending on the direction of the transfered wave number. In this case due to the anisotropic position of
the Dirac points in the Brillouin zone, each of the Dirac points has not identical contribution
for this type of the quantities, even when the Dirac point approximation is valid. Meanwhile, since the inter-valley transitions could take place between the different Dirac cones, these type of transitions cannot be captured within the single Dirac cone approximation.

Anisotropic Friedel oscillations in two-Dimensional structures have been observed before
\cite{hofmann1997anisotropic}. However, in the present case the anisotropic effects
are direct manifestation of non-identical contribution of Fermi circles and Dirac points
in wave number dependent quantities. Some of the physical quantities, such as dielectric
function, are given by integration over the Brillouin zone as expressed in Eq. (\ref{aa}).
This integration goes beyond the states in which the Dirac point approximation and the linear
dispersion relation no longer valid. However distribution function at low temperatures and
Fermi energies picks up the contribution of the states which were located near to the
Dirac points. Nevertheless for evaluating non-scalar quantities that depending on the
wave number and its direction we should consider that the orientation of the wave number, $q$,
(relative to the position vector of the Dirac points in k-space) determines the contribution
of each Dirac point. It can be clearly demonstrated that the dielectric function shows
anisotropic directional dependence in $q$-space.

It should be noted that increasing of the Fermi energy results in deformation of
circle-like Fermi curves (Fermi-circles) of low Fermi energies around the Dirac points \cite{Shih2017}.
In this case the isotropic form of the Fermi circles change into the trigonal-shaped
contours (known as trigonal warping effect) and the isotropic form of the Fermi curve
around of the Dirac points has totally been removed at high Fermi energies. This type
of deformation could results in a new source of anisotropy at high Fermi energies which could be captured within the single cone approximation. Meanwhile trigonal warping induced anisotropy
has not been considered in the current work which was limited to the low Fermi energies.

At low Fermi energies optical transitions around each Dirac points, where take
place within a single Fermi circle, have the main contribution in the dielectric
function of the honeycomb systems. This manifests itself as a central
peak of the dielectric function in the middle of the Brillouin zone. It could be
shown that beyond the Dirac point approximation the central peak contains the
contribution of all of the Dirac points via the inter-band transitions.

When the Fermi energy is close to Dirac points ($E_F=0$) and at the high
wave length limit ($q\ll \mid K_D-K'_D $) the occupation factor $f^s_k-f^{s'}_{k'}$
could have a significant value only when the $k$ and ${k'}$ states are close to a single Dirac
point of different bands ($s\neq s'$). Meanwhile the Kronecker delta, $\delta_{\vec{k'},\vec{k}+\vec{q}}$,
in the expression of the polarization indicates that the contribution of the Dirac
points should be selected by $\Gamma$-point $(\vec{q}=0)$. Since at this limit the
main contribution is due to the intra-valley transitions which take place near
the Dirac points in which $k\approx K_D$ and $k'\approx K_D$ therefore the
mentioned Kronecker delta, which reflects the momentum conservation,
imposes that the contribution of the Dirac points should be manifest themselves
at the $\Gamma$-point  $(\vec{q}  \approx 0)$ of the $q$-space.

\begin{figure}[t]
	\includegraphics[width=0.5\linewidth]{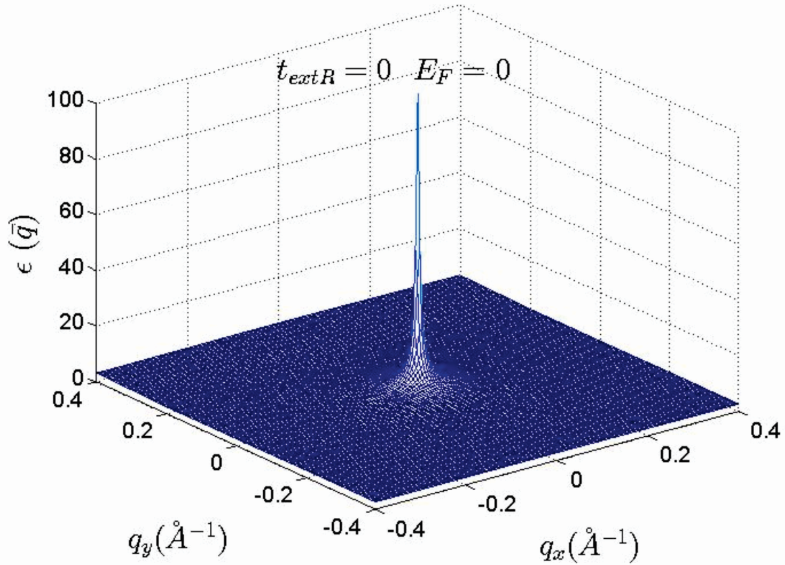}
	\centering
	\caption{$k$-space dielectric function of monolayer graphene at $t_{extR}=0$.
		Additionally it was assumed that the intrinsic spin-orbit coupling has been
		also negligible. This enables us to compute the net band induced anisotropic
		effects. \label{eps-mg-0}}
\end{figure}

\begin{figure}[t]
	\includegraphics[width=0.75\linewidth]{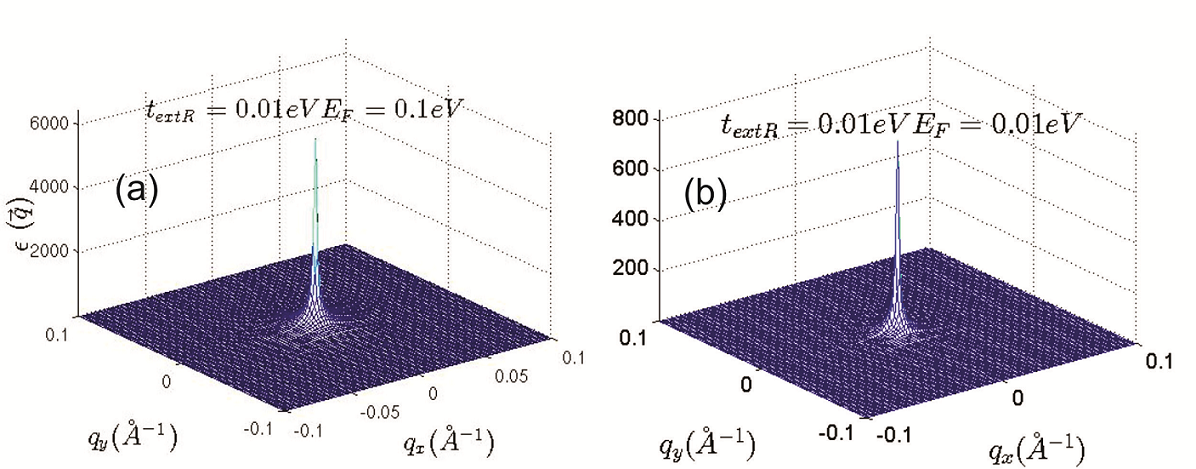}
	\centering
	\caption{$k$-space dielectric function of monolayer silicene and germanene $t_{extR}=0.01eV$.
		\label{eps-SG}}
\end{figure}

The mentioned argument reveals the fact that the central peak (Figs. \ref{eps-mg-0} and \ref{eps-SG}) of the dielectric
function is exactly sum of all of the intra-valley contributions from each Dirac point. In
this case since the main contribution belongs to the case of zero transferred
momentum ($q\sim 0$) the question of the anisotropy in $q$-space seems to be
irrelevant at $E_F=0$. Although the contribution of the intra-valley transitions are dominant at $q\sim 0$,  however, this is not the case for the wave numbers within the range of $ \mid K_D -K'_D \mid-2k_F \leq q< \mid K_D -K'_D \mid+2k_F$. It should be noted that
the contribution of inter-valley transitions between different Dirac points ($q\sim \mid K_D -K'_D \mid$)
are not identical. In this case the contribution of each pair of contributing Dirac point (${\bf{K}}_D$ and ${\bf{K}'}_D$) has been determined by the direction of the $\vec{q}$ vector.

\begin{figure}[h]
	\includegraphics[width=0.7\linewidth]{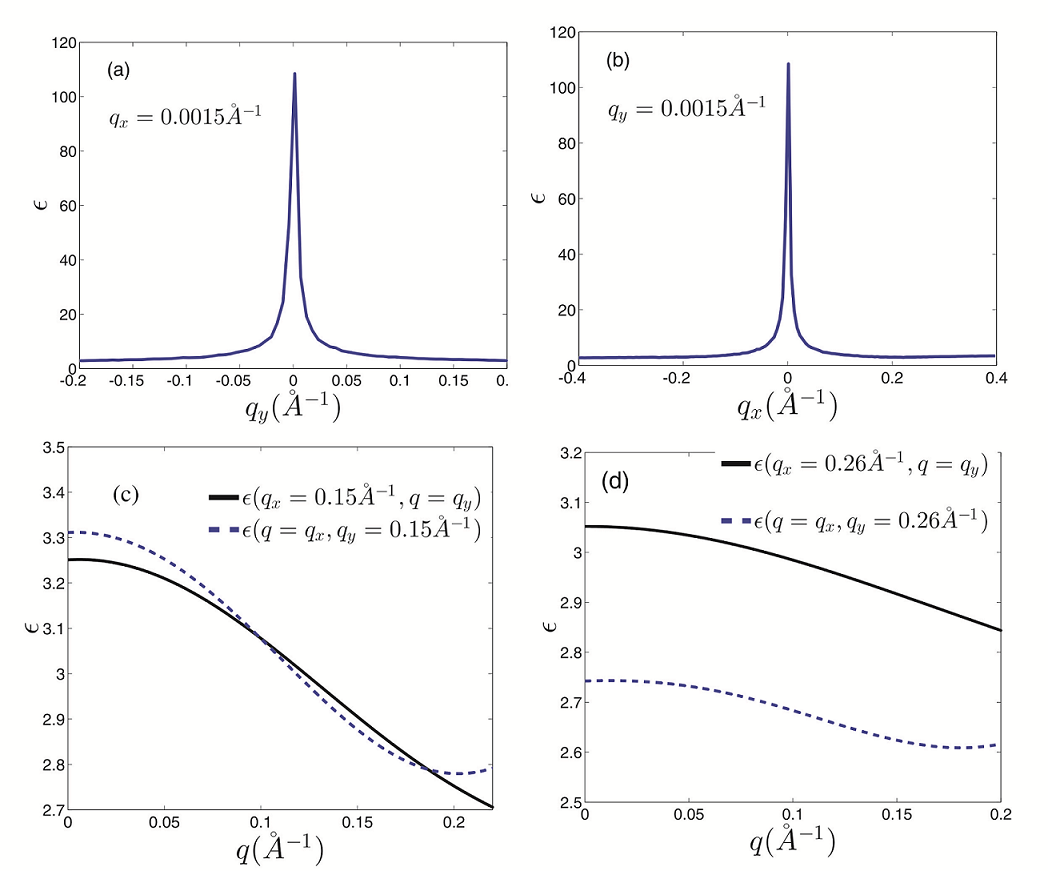}
	\centering
	\caption{Dielectric function of monolayer graphene at different symmetrically
		chosen slices for $t_{extR}=0$ in $k$-space. (a) At $q_x=0.0015{\AA}^{-1}$ plane.
		(b) $q_y=0.0015{\AA}^{-1}$ plane. (c) At $q_x=0.15{\AA}^{-1}$ and $q_y=0.15{\AA}^{-1}$
		slices. (d) At $q_x=0.26{\AA}^{-1}$ and $q_y=0.26{\AA}^{-1}$ planes. Anisotropic effects
		appear far away from the origin where the contribution of the inter-valley transitions
		should be taken into account.\label{eps-mg-0-2d}}
\end{figure}

\begin{figure}[t]
	\includegraphics[width=0.8\linewidth]{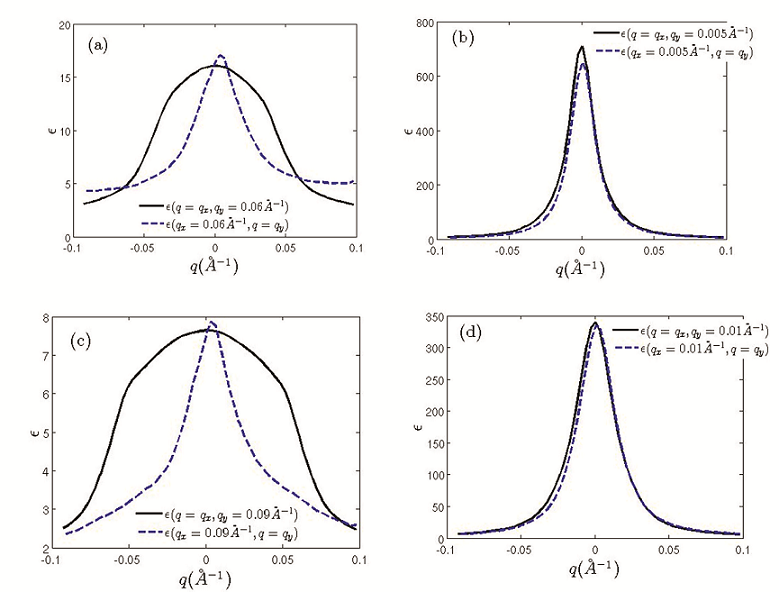}
	\centering
	\caption{Dielectric function of monolayer silicene at different symmetrically
		chosen slices for $t_{extR}=0.01 $eV in $k$-space. (a) At $q_\alpha=0.06{\AA}^{-1}$ planes ($\alpha = x,y$).
		(b) $q_\alpha=0.005{\AA}^{-1}$ planes. (c) At $q_\alpha=0.09{\AA}^{-1}$ 
		slices. (d) At $q_\alpha=0.01{\AA}^{-1}$ planes. Anisotropic effects
		appear far away from the origin where the contribution of the inter-valley transitions
		is dominant.\label{eps-s}}
\end{figure}

\begin{figure}[t]
	\includegraphics[width=0.8\linewidth]{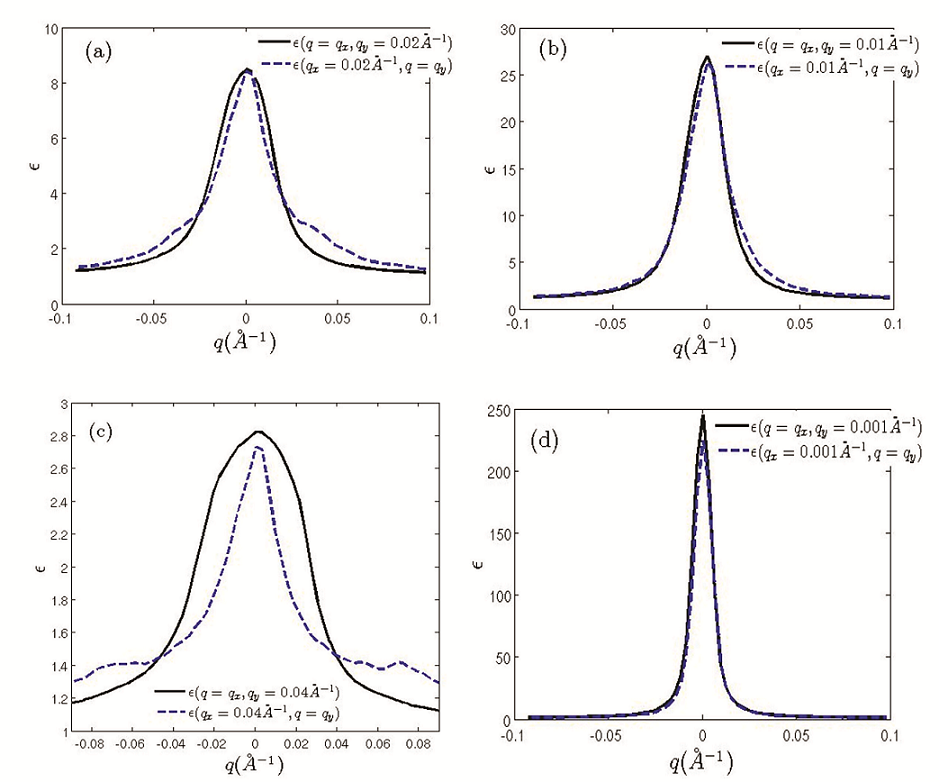}
	\centering
	\caption{Dielectric function of monolayer silicene at different symmetrically
		chosen slices for $t_{extR}=0.01 $eV in $k$-space. (a) At $q_\alpha=0.02{\AA}^{-1}$ planes ($\alpha = x,y$).
		(b) $q_\alpha=0.01{\AA}^{-1}$ planes. (c) At $q_\alpha=0.04{\AA}^{-1}$ 
		slices. (d) At $q_\alpha=0.001{\AA}^{-1}$ planes. Anisotropic effects
		appear far away from the origin where the contribution of the inter-valley transitions
		is dominant.\label{eps-g}}
\end{figure}

In order to obtain the anisotropic effects which have been induced
merely by band energy we have to switch off both types of the spin
orbit couplings. It was reported that the Rashba coupling strength
in graphene is greater than the intrinsic spin orbit interaction \cite{dedkov2008rashba}.
Therefore one can ignore intrinsic spin orbit
coupling in monolayer graphene. At zero
Rashba interaction both intrinsic and extrinsic spin-orbit couplings
are absent. This enables us to obtain the anisotropic effects which
could be induced merely by band energy. Dielectric function at zero
Rashba coupling and zero Fermi energy ($E_F=0$) has been obtained as
depicted in Fig. (\ref{eps-mg-0}). At the first look it seems that
there is no anisotropy in the dielectric function of the honeycomb structures (Figs. \ref{eps-mg-0} and \ref{eps-SG}), however, it should be noted that the anisotropy
of the dielectric function has been hidden behind the large central
peak at $q\sim 0$. The amount of the dielectric anisotropy is
very small in comparison with the value of the dielectric function at
the $\Gamma$-point. Accordingly this fact prevents the identification of the
directional dependence of the dielectric function.  At low wave
numbers i.e. in the range of the intra-valley transitions dielectric
function seems to be quite isotropic in $q$-space Fig. (\ref{eps-mg-0-2d}
(a) and (b)). However, far from the central region if we select different
symmetric slices of the dielectric surface, the anisotropy of the dielectric
function will be evident Fig. (\ref{eps-mg-0-2d} (c) and (d)). A similar discussion holds for the case of silicene and germanene where as shown in Figs. \ref{eps-s} and \ref{eps-g} dielectric function has completely different behavior on symmetrically chosen slices, especially far enough the $\Gamma$ point. These figures evidently  
show that for different directions in the $k$-space behavior of the
dielectric function are quite different. This reveals that the contributing
inter-valley transitions in finite wave length limit ($q\sim \mid K_D -K'_D \mid$)
introduce the anisotropic behaviors.

\begin{figure}[h]
	\includegraphics[width=0.5\linewidth]{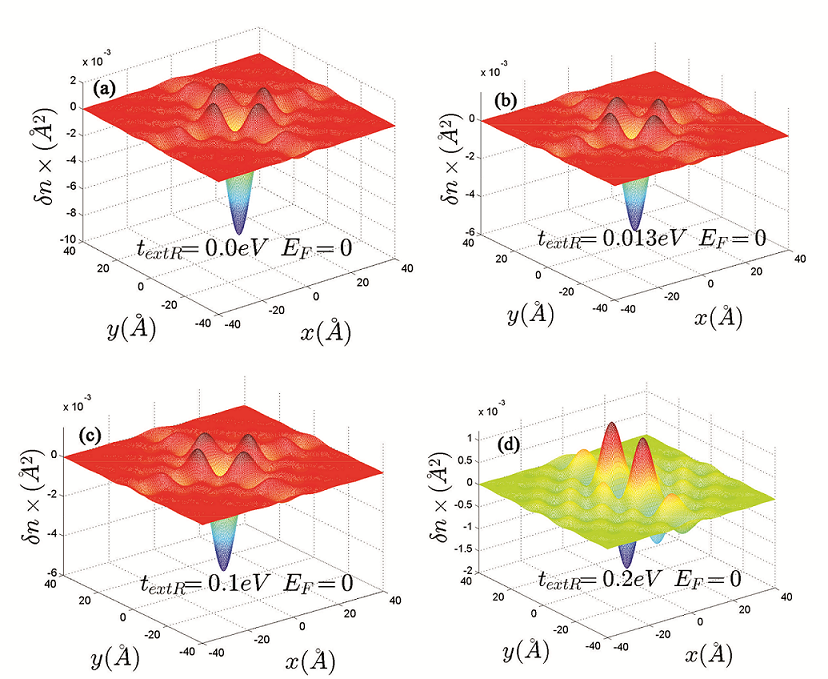}
	\centering
	\caption{Real space anisotropic Friedel oscillations in monolayer
		graphene at different Rashba couplings. (a)-(c) As shown in these figures the Rashba interaction has not a significant influence on the Friedel oscillations at intermediate Rashba coupling strength. (d) Unlike to the previous case for $t_{extR}=0.2$~eV Friedel oscillations are not similar even for $x$ and $y$ directions. \label{Fr-mg}}
\end{figure}
\begin{figure}[h]
	\includegraphics[width=0.5\linewidth]{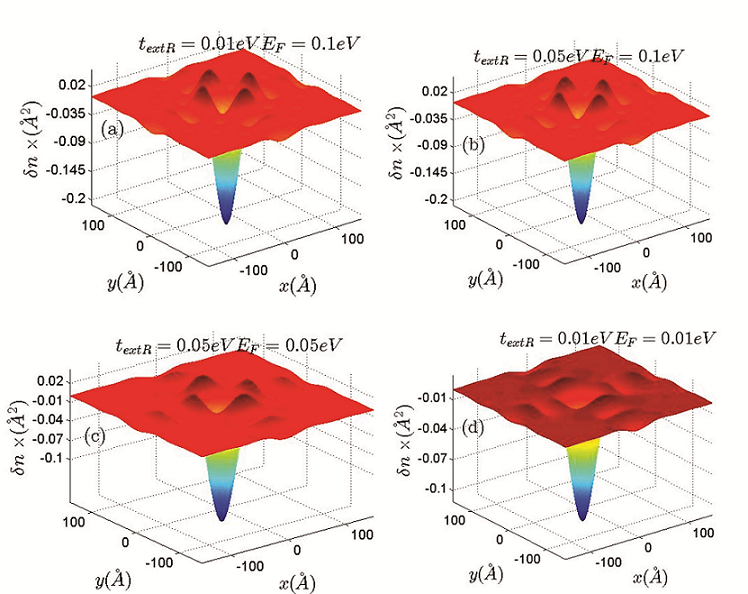}
	\centering
	\caption{Real space anisotropic Friedel oscillations in monolayer
		silicene ((a) and (b)) and germanene ((c) and (d)) at different Rashba couplings. The Rashba interaction has not a significant influence on the Friedel oscillations. \label{Fr-s}}
\end{figure}
\begin{figure}[t]
	\includegraphics[width=1.0\linewidth]{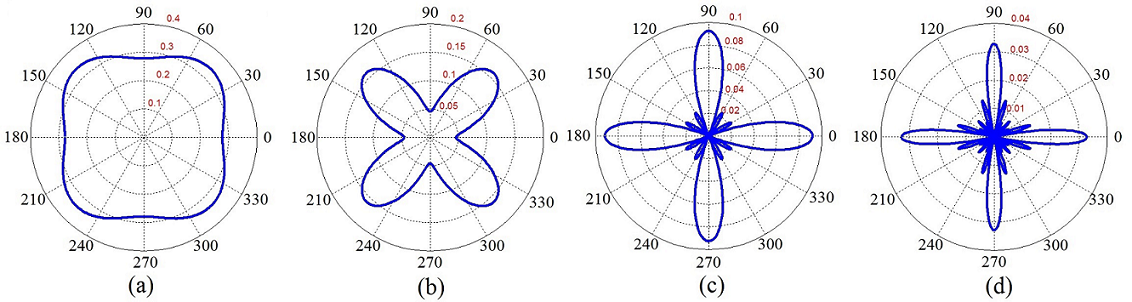}
	\centering
	\caption{Angular dependence of the normalized graphene Friedel oscillations, $\delta n(r,\theta)/n_0$, for $t_{extR}=0$, $E_F=0.1 t$ and $K_BT=0$ at (a) $r=4\AA$, (b) $r=5\AA$, (c) $r=14\AA$ and (d) $r=25\AA$ in which we have defined $n_0=\delta n(\textbf{r}=0)$.  \label{polar}}
\end{figure}
As discussed before   band induced anisotropic effects ( see Fig.~\ref{Dirac1}) have
been reflected in the Friedel oscillations of the graphene-like structures illustrated in Figs. \ref{Fr-mg} and \ref{Fr-s}.
The anisotropy of the dielectric function as discussed before is due to the non-identical contribution
of the Dirac points and the nonlinear part of the spectrum cannot contribute
in dielectric function when the Fermi level is close to the Dirac points.
Increasing the Rashba coupling strength slightly modifies the Friedel oscillations in silicene, however the change in monolayer graphene is relatively more
pronounced compared to that of other selected honeycomb structures
where Figs. (\ref{Fr-mg} and \ref{Fr-s}) exactly indicates this fact.
This could be explained if we consider relatively large and dominant
intrinsic spin-orbit coupling in silicene.

As shown in the figure (\ref{Fr-mg}) the anisotropic
Friedel oscillations have been observed even when the Rashba coupling
strength is very low or zero. It can be inferred from the
results of the current work that the Rashba coupling is less effective
in the generation of the anisotropy. Therefore one can conclude that
the anisotropy of the dielectric function and Friedel oscillations
mainly depends on the anisotropy of the band structure in $k$-space.

There are several studies which have been performed in this field, aiming at an accurate
quantitative prediction of dynamical dielectric function, screened charged impurity
potential and Friedel oscillations in graphene-like materials. It was realized that
the long-distance decay of Friedel oscillations in graphene depends on the symmetry
of the scatterer\cite{CheianovEPJ}. In addition a faster, $\delta n \sim 1/r^3$,
decay in comparison with conventional 2D electron systems has been observed in
Friedel oscillations of a localized impurity inside the monolayer graphene within
the Dirac point approximation \cite{CheianovEPJ,wunsch2006dynamical}. However,
$1/r$ decay has been reported for bilayer graphene \cite{BenaPRL} and strong
asymmetry and an inverse square-root decay has also been obtained for an
anisotropic graphene-like structure when one of the nearest-neighbor hopping
amplitudes is different from the others \cite{dutreix2013friedel}. Recently in rhombohedral graphene multi-layers, $1/r$ decay has been observed for impurity induced Friedel oscillations \cite{Dutreix2016}. Completely
isotropic behavior has been reported for the potential of a screened charged
impurity, Friedel oscillations \cite{scholz2012dielectric,wunsch2006dynamical,gomez2011measurable,Cheianov}
and static dielectric function \cite{hwang2007dielectric} within the Dirac
point approximation in graphene. Similarly the Dirac point approximation
results in isotropic screened potential of a charged impurity in other
graphene-like materials such as silicene and germanene \cite{tabert2014dynamical}.

The studies based on Dirac point approximation give
the correct physics of the high wave length limit ($q << k_F$) at $E_F=0$ where inter-valley
transitions could not contribute in the physical processes.
In the absence of the spin-orbit couplings by using the massless linear
Dirac spectrum it was also shown that short wavelength spatial dependence
of the local density of states leads to anisotropic Friedel oscillations which has the form\cite{Attila}
\begin{equation}\label{ff}
{{ \delta n}}\left( r \right)\sim c(\vec{r}) \rho_0(E_F)\frac{\sin(2k_Fr)}{r^2}.
\end{equation}
In which $c(\vec{r})$ is the short wavelength spatial dependence factor and $\rho_0(E)$
is the density of states. Anisotropic dependence of the Friedel oscillations has been
introduced by $c(\vec{r})$ factor which was found to be invariant
under threefold rotations\cite{Attila}. However, if the impurity could not produce inter-valley
scatterings this factor is reduced to a constant number\cite{Attila}. Therefore the anisotropic
effects have been removed in the absence of inter-valley transitions\cite{Attila}.
In the current study we have observed that for finite Fermi energies $0<E_F \leq 1$eV even
intra-valley transitions are the source of the anisotropic behaviors at linear energy dispersion regime.

In the case of the single valley band structures where all of the transitions should be considered as intra-valley transitions. The wave length of the Friedel oscillations is modulated by Fermi wave number. However it can be easily shown that this is not the case for multi-valley band structures. In which the inter-valley transitions could contribute in the dielectric function.  
As indicated in Eq. (\ref{ff}) it was expected that the wavelength of the Friedel oscillations should be modulated by the Fermi wave-vector $k_F$ \cite{Attila}.  Where the long range behavior of the local density of states has been obtained within the single valley approximation and linear dispersion relation\cite{Attila}. The possible transfered momentums, $q$, determine the oscillation wavelength of the induced charged and for single valley band structures in two-dimensional systems typical transfered momentum is $q\sim2k_F$. 
However it should be noted that for a typical graphene Fermi energy e.g. $E_F=0.1$eV one can obtain $k_F=E_F/(\hbar v_F)=0.0152\AA^{-1}$. Therefore the oscillation wavelength that corresponds to the intra-valley transitions is about $\lambda_{intra}=2\pi/2k_F\sim200\AA$. On the other hand in the present case the inter-valley transitions with momentum transfer of $q\sim|\textbf{K}-\textbf{K}'|\pm 2 k_F$  correspond to the oscillations with wavelength of $\lambda_{inter}=2\pi/(|\textbf{K}-\textbf{K}'|\pm 2 k_F)$. The wavelengths of the oscillations in the current work for monolayer graphene are in the following range $7\AA\lesssim\lambda_{inter}\lesssim13\AA$  that are at the same order of the inter-valley transition wavelengths given by $\lambda_{inter}=2\pi/(|\textbf{K}-\textbf{K}'|\pm 2 k_F)$. For example the distance between two successive Dirac points in graphene is about $\Delta K=|K-K'| \sim 1.7\AA^{-1}$. Therefore the average momentum transfer between these Dirac points is $ q \sim |K_-K'|/2=0.85\AA^{-1}$ then  $\lambda_{inter}=2\pi/\Delta K=7.37\AA$. 

The numerical results indicate that the wavelength of the oscillations is less-sensitive to the value of the Fermi energy. This can be realized  if we consider that  $|\textbf{K}-\textbf{K}'|>>2k_F$ for intermediate Fermi energies.

Decay rate of the Friedel oscillations are determined by fitting to the numerical results. We have examined several decay rates such as $1/r$, $1/r^2$ and $1/r^3$. Numerical fitting shows that the $1/r$  decay rate is much more close to the computational data profile. More precisely decay rate is actually $1/r^{1+\eta}$ where $0<\eta<0.2$. 

Another important issue about the Friedel oscillations is that how sharp the mentioned density anisotropy really is? In this way we have obtained the angular dependence of the induced density at different distances as depicted in Fig. (\ref{polar}). As indicated in this figure the anisotropy of the Friedel oscillations increases by distance. It  can be realized that the angular dependence of the induced density is so sharp at intermediate distances. This provides more detectable condition for observation of the anisotropy. 

Interestingly it was shown that the Friedel oscillations in graphene have a strong sublattice
asymmetry\cite{lawlor}. These calculations have been performed beyond the Dirac point
approximation within the Born approximation which can be employed for weak scattering
potentials and the stationary phase approximation (SPA) has also been applied for Brillouin
zone integrations\cite{lawlor}. Anisotropic Friedel oscillations could also be inferred
from the numerical results of the recent work in the absence of the spin-orbit interactions
especially over short distances.

Finally it is important to note, the anisotropy of the dielectric
function suggests that the orientation of the bases vectors of the
honeycomb lattice could be determined by full optical measurements.
Since dynamical dielectric function of the graphene-like materials
are possibly have the same anisotropic nature. Therefore the absorption
spectra of the system should be anisotropic. Accordingly, the real space
orientation of the basis vectors could be explored since the absorption spectra
leads to identification of the band energy configuration in k-space.

\bibliography{refrences}



\section*{Author contributions statement}

All authors contributed to the numerical programming. A. Phirouznia and T. Farajollahpour carried out the discussion and analysed the results. In addition final version of the manuscript has been reviewed by A. Phirouznia and T. Farajollahpour.  

\section*{Additional information}

Competing financial interests: The authors declare no competing financial interests.

\end{document}